\documentclass[12pt,preprint]{aastex}

\shorttitle{The updated DiFX software correlator}
\shortauthors{Deller, Brisken, Phillips, Morgan, Alef, Cappallo, Middelberg, Romney, Rottman, Tingay \& Wayth}

\newcommand{\degrees}{\ensuremath{\,^{\circ}}}

\begin{document}

\title{DiFX2: A more flexible, efficient, robust and powerful software correlator}

\author{A.T. Deller\altaffilmark{1,2}, W.F. Brisken\altaffilmark{1}, C.J. Phillips\altaffilmark{3}, J. Morgan\altaffilmark{4}, W. Alef\altaffilmark{5}, R. Cappallo\altaffilmark{6}, E. Middelberg\altaffilmark{7}, J. Romney\altaffilmark{1}, H. Rottmann\altaffilmark{5},  S.J. Tingay\altaffilmark{4} \& R. Wayth\altaffilmark{4}}
\altaffiltext{1}{National Radio Astronomy Observatory, PO Box O, Socorro, NM, 87801, USA}
\altaffiltext{2}{UC Berkeley, 601 Campbell Hall, University of California at Berkeley, Berkeley, CA, 94720, USA}
\altaffiltext{3}{Australia Telescope National Facility, CSIRO Astronomy \& Space Sciences, PO Box 76, Epping, NSW, Australia}
\altaffiltext{4}{International Centre for Radio Astronomy Research, Curtin University, GPO Box U1987, Perth, WA, Australia}
\altaffiltext{5}{Max--Plank--Institut f\"{u}r Radioastronomie, Postfach 20 24, 53010 Bonn, Germany}
\altaffiltext{6}{MIT Haystack Observatory, Westford, MA, 01886, USA}
\altaffiltext{7}{Ruhr-Universit\"{a}t Bochum, Astronomisches Institut, NA 7/73, Universit\"{a}tstra. 150, D-44801 Bochum, Germany}

\begin{abstract}
Software correlation, where a correlation algorithm written in a high--level language such as C++ is run on commodity computer hardware, has become increasingly attractive for small to medium sized and/or bandwidth constrained radio interferometers.  In particular, many long baseline arrays (which typically have fewer than 20 elements and are restricted in observing bandwidth by costly recording hardware and media) have utilized software correlators for rapid, cost--effective correlator upgrades to allow compatibility with new, wider bandwidth recording systems and improve correlator flexibility.  The DiFX correlator, made publicly available in 2007, has been a popular choice in such upgrades and is now used for production correlation by a number of observatories and research groups worldwide.  Here we describe the evolution in the capabilities of the DiFX correlator over the past three years, including a number of new capabilities, substantial performance improvements, and a large amount of supporting infrastructure to ease use of the code.  New capabilities include the ability to correlate a large number of phase centers in a single correlation pass, the extraction of phase calibration tones, correlation of disparate but overlapping sub--bands, the production of rapidly sampled filterbank and kurtosis data at minimal cost, and many more.  The latest version of the code is at least 15\% faster than the original, and in certain situations many times this value.  Finally, we also present detailed test results validating the correctness of the new code.
\end{abstract}


\section{Introduction}
Development of the Distributed FX (DiFX) software correlator began in 2005, primarily for usage with the Australian Long Baseline Array (LBA) as part of a sensitivity upgrade program \citep{deller07a}, where it entered production usage in 2006.  It is an FX style correlator \citep[see e.g.,][]{chikada87a,thompson94a,romney99a} designed to run on modern CPUs under Linux or Mac OS X.  The basic principles of radio astronomy cross--correlator fundamentals will not be re--derived in this article, which focuses on the particular implementation of the DiFX software correlator.  We direct the reader to the references above for a thorough explanation of FX--style correlator functionality, and to \citet{deller07a} for a comprehensive description of the specific implementation of this functionality in the DiFX code.

The DiFX code is accelerated using vector arithmetic libraries, specifically the Intel Performace Primitive library\footnote{software.intel.com/en-us/intel-ipp/}, and the distribution across multiple nodes is enabled using implementations of the Message Passing Interface\footnote{http://www.mcs.anl.gov/research/projects/mpi/}. The advantages brought by the adoption of a newer, more flexible correlator architecture were enumerated by \citet{deller07a}, and included greater flexibility in the setting of correlation parameters, lower cost, rapid development, ease of maintenance, and upgradeability (both in hardware and software).  In keeping with this final point, development of the DiFX software correlator has continued rapidly since its first public release in 2007.  Since that time, many new features and performance improvements have been merged into the DiFX codebase, which cumulatively are sufficient to merit a major version increment for the DiFX package, which we designate DiFX2.  Where necessary, any release of DiFX prior to DiFX2 will be referred to generically as DiFX1.x, acknowledging that this time period spanned a series of release numbers.  Some of the features presented here were made available early in the DiFX1.x series and have been available for several years, while others were added recently and are only available in DiFX2.

DiFX has been adopted by a number of leading VLBI facilities in addition to the LBA.  Specifically, the Very Long Baseline Array (VLBA) operated by the National Radio Astronomy Observatory (NRAO\footnote{The National Radio Astronomy Observatory is a facility of the National Science Foundation operated under cooperative agreement by Associated Universities, Inc.}) in the US has retired its hardware correlator, which was designed in the 1980s, and migrated completely to DiFX. In addition, the Max Plank Institute for Radioastronomy (MPIfR) has begun routine operation of DiFX in parallel with the existing Mark4 hardware correlator operations, and plans to phase out its Mark4 hardware correlator by the end of 2010.  The specific needs of the LBA, VLBA and MPIfR have driven the development of many of the new capabilities of DiFX, including the features discussed below which allow entirely new areas of long--baseline science to be undertaken.  All of these developments have been made available to all current and potential users.

Almost as important for new users of DiFX, considerable effort has been made to improve the documentation and online resources available for installing and testing DiFX.  A collaboratively managed wiki is available at \verb+http://cira.ivec.org/dokuwiki/doku.php/difx/start+, and two mailing lists are available to seek or disseminate information regarding DiFX.  One list reaches the entire DiFX community, while the second is focused specifically on code developers.  Prospective users of DiFX are directed to the wiki, where further information is available on how to obtain the code from the central repository.  Finally, improved version control has been implemented since the early days of DiFX1.x, and tagged releases of frozen code are made available on a regular basis.  The current stable version of DiFX2 is DiFX-2.0.0, and the current version of the previous series is DiFX-1.5.4.

In this article, we describe the new DiFX2 functionality in \S\ref{sec:newfeatures}.  The performance improvements which have been provided in the accompanying code changes are listed and quantified in \S\ref{sec:performance}.  Additional code which provides functionality for DiFX2 that is not directly part of the correlator itself is described in \S\ref{sec:operations}, and the validation testing undertaken for DiFX2 is described in \S\ref{sec:validation}.  Future work is discussed in \S\ref{sec:futurework}, and our conclusions are presented in \S\ref{sec:conclusions}.

\section{New features}
\label{sec:newfeatures}
\subsection{FITS-IDI and Mark4 format output}
\label{subsec:fitsidi}
Initially, DiFX1.x only supported the RPFITS\footnote{http://www.atnf.csiro.au/computing/software/rpfits.html} file format, which was the format historically used by the LBA.  However, RPFITS is not a standard FITS format and has limited support in most radio astronomy post-processing packages. Accordingly, in version 1.5.0 the ability to produce FITS-IDI format correlator files (as produced by the VLBA and the Joint Institute for VLBI in Europe [JIVE] hardware correlators) was added to DiFX.  Unlike the RPFITS files written by early versions of DiFX1.x, the FITS file is not directly written by the correlator, but is translated from a ``DiFX" binary output written by DiFX2 after correlation has completed.  This eliminates the need to link large FITS libraries into DiFX, and simplifies and speeds the output writing process.  Support for the RPFITS format has been withdrawn in DiFX2.

Geodetic observers typically use specialized post--processing software such as HOPS\footnote{see http://www.haystack.mit.edu/tech/vlbi/hops.html} which is closely tied to the visibility data format produced by the Mark4 hardware correlator \citep{whitney04a}.  In order to facilitate the use of DiFX2 for geodetic observations, an additional translation program has been written to produce these Mark4 format visibility datasets from the binary DiFX output.  The ability to import DiFX format output data directly into the HOPS geodetic post--processing package is new in DiFX2.

\subsection{Native Mark5 interface}
\label{subsec:nativemk5}
The Mark5 recording media series \citep{whitney03a} is widely used amongst VLBI networks, with the VLBA, the European VLBI Network (EVN), the Korean VLBI Network (KVN) and global geodetic arrays all utilizing the system.  Initially, however, DiFX1.x could not read data directly from Mark5 disk modules, as it required the data to be accessible from a standard Linux file. Correlating Mark5 data therefore required a tedious intermediate step of exporting the files to a Linux filesystem, which imposed additional overhead in time and storage space. Since version 1.5.0, DiFX has had the ability to read Mark5 disk modules ``natively", using the application interface available from the Mark5 vendor.  This eliminates the need to export data from modules to standard files, streamlining the correlation process and reducing the need for large amounts of standard disk storage.  The support of VLBA and Mark4 data formats on both standard Linux files and Mark5 modules has been extended to include Mark5B, and limited support is already in place for the next--generation VDIF format described by \citet{whitney09a}.

\subsection{Phase calibration tone extraction}
\label{subsec:phasecal}
Phase calibration tones can be injected at the front end of a radio astronomy antenna in order to provide a convenient means to estimate instrumental delays.  For applications such as geodesy, phase calibration tones are heavily relied upon.  Whilst some radiotelescope arrays extract, average and store phase calibration tone information at the antenna, others rely on the correlator to perform this important function.  Accordingly, a flexible phase calibration tone extraction system has been added to DiFX2.  The phase calibration extraction in DiFX2 can be configured to extract any number of tones, unlike many existing hardware implementations such as those at the VLBA stations (which provide two tones per sub--band).  Preliminary results comparing DiFX corrections to those extracted at the VLBA stations show agreement to $\ll1$\degrees\ (corresponding to $\sim$ femtoseconds at 43 GHz), and also show that the computational overhead of extracting all the phase calibration tones present\footnote{which are spaced at 1 MHz intervals, so typically 8 or 16 tones per sub--band} is $\sim$5\%.  A detailed analysis of the performance of DiFX on geodetic observations, including the verification of DiFX2 phase calibration extraction and the production of Mark4 format visibility data, is deferred to a future publication (Morgan et al., in preparation).

\subsection{Spectral selection and averaging}
\label{subsec:specselectavg}
Once the data has been channelized (the ``F" portion of the FX algorithm) it is possible to discard segments of the spectrum which hold no interest for the current observation.  The main application of such ``spectral selection" is to zoom in on widely separated spectral features such as masers which are contained within a wide bandwidth.  Use of this new feature in DiFX2, which we generically term ``zoom mode", reduces the load on the cross-multiply/accumulate (``X") portion of the FX algorithm, and more importantly reduces the amount of data which must be returned to the manager node for long-term accumulation \citep[see ][]{deller07a}.  This allows very high spectral resolution to be obtained without overloading the correlator interconnect, and without generating unduly large amounts of data to be written to intermediate disk results and later discarded (as was the case with all versions of DiFX1.x).

An alternate application of spectral selection, which we generically term ``band--matching", facilitates the correlation of heterogeneous recorded bands by subdividing wider bands recorded at some antennas to match narrower bands recorded at other antennas.  This is a particularly useful feature for correlating infrequently used antennas with non-standard VLBI backend systems, which are unable to produce bands compatible with other VLBI systems.  A current example involves geodetic correlations where 32 MHz bands recorded at most stations are correlated against 16 MHz bands recorded at the Plateau de Bure interferometer.  The correlation of upper sideband data with lower sideband data (covering the same spectral range) is also supported.

Spectral averaging allows multiple spectral points to be averaged after correlation, but before the visibilities are returned to the manager node.  This is useful in two cases.  The first is when the desired final spectral resolution is low.  In this instance, generating such coarse spectral resolution directly by means of a very short Fourier transform is not efficient, so use of an optimally sized transform (typically of length $\sim$256 points) is preferred, with spectral averaging before the visibilities are returned to the manager.  As with spectral selection, this saves correlator interconnect and disk space resources.

The second, and more important, usage of spectral averaging is with the multiple phase center correlations described in the following section.  In order to avoid bandwidth decorrelation effects as described below, very high spectral resolution is required initially.  After the visibilities have been shifted to the desired phase center, however, they can be averaged to a standard VLBI resolution.  Spectral averaging is critical for this application, as the return of multiple copies of very high spectral resolution visibilities would completely overwhelm the correlator interconnect.  

\subsection{Multiple simultaneous phase centers}
\label{subsec:mpc}
Because of the very high fringe rates inherent in VLBI observations, the use of standard frequency (100s kHz) and time resolution (seconds) leads to an extremely small (several arcsecond) field of view \citep[see e.g.,][]{middelberg11a}. We hereafter refer to a narrow field of view resulting from standard VLBI correlation parameters as a ``pencil beam".  \citet{thompson94a} contains a detailed explanation of the challenges inherent in wide--field imaging.  Whilst improving the temporal and spectral resolution allows somewhat wider fields of view (and was indeed one of the main drivers of DiFX development), this carries an increasingly problematic cost in expanded data volume.  Mapping even a tenth of the primary beam of the VLBA at 1.4 GHz with no more than 10\% decorrelation due to time and bandwidth decorrelation requires 8 kHz frequency channels and an integration time of 0.1s -- which yields visibility datasets $>$2 TB for a typical 12 hour VLBA observation at current bandwidths.

An alternative to mapping large swathes of sky (which in any case are almost entirely empty at VLBI resolution at cm frequencies) is to image small areas around known sources.  This can be accomplished by shifting the phase center of the correlation (a ``$uv$ shift") to the location of known sources and averaging visibilities to obtain manageable--sized datasets, which can be used to produce pencil beams at new locations \citep[see e.g.,][]{lenc08a}.  A $uv$ shift is implemented by calculating the baseline--based differential geometric delay between the desired and applied phase center, converting to a phase rotation for each visibility by multiplying this delay by the associated sky frequency, and rotating the visibility phases by this value.  \citet{morgan10a} examine the problem of $uv$ shifting in more detail, including the detailed calculation of the necessary delay shifts.  The drawback of using this approach after correlation is the necessity of generating an initial visibility dataset which is as large as that required for a single large image. The intermediate data volume problem is therefore comparable to that experienced with the single large image approach, and the I/O cost of writing visibilities to and reading from disk is substantial.

If implemented within the correlator, however, the twin problems of I/O and storage volume are solved, because the intermediate data products (the high spectral resolution visibilities held at the processing nodes) do not need to be transmitted from where they are calculated, and likewise are not written to disk.  Obtaining sufficiently high time resolution is trivially implemented -- the time division multiplexing within DiFX \citep[see ][]{deller07a} already provides time resolution better than that required in most cases, but the ability to $uv$ shift and average after a shorter, user--specified time has also been implemented.  For $P$ phase centers, the processing nodes transmit $P$ normal--sized (post--average) collections of visibility results back to the manager node, and $P$ normal--size datasets are ultimately written to disk.  The impact on performance of this feature is relatively small, due to the fact that the $uv$ shift/average operations need only be carried out relatively infrequently (compared to the multiplications, additions, and Fourier transforms required for the regular correlation process).   

The visibility amplitudes and weights are corrected for time and bandwidth decorrelation on--line, before the visibilities are written to disk.  The resultant $P$ pencil beams can be reduced and imaged using standard tools.  Presently, these corrections are not tabulated and saved, since no post--processing software exists which could parse and use this information.  The information is readily available, however, and could be formatted and written out when suitable post--processing becomes available.

\begin{figure}
\includegraphics[width=0.9\textwidth]{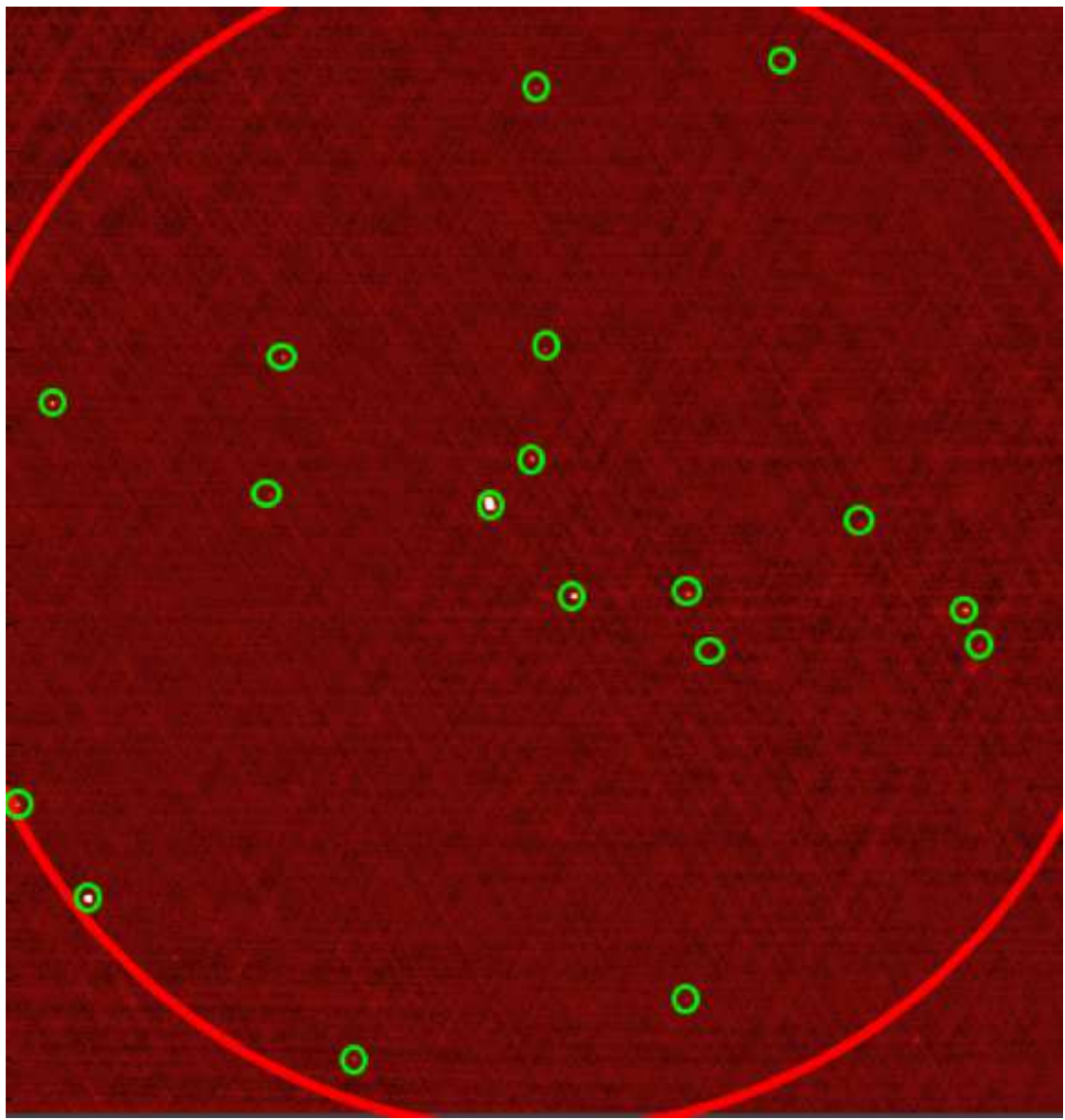}
\caption{Example of a representative finding field centered on 07h45m07.270s +33\degrees40'37.52" (from the FIRST survey -- http://sundog.stsci.edu/).  The bold black ring shows the 31' primary beam of a 25m dish at 1600 MHz, and the small white rings show the individual pencil beams that would be placed on known sources.  The pencil beam diameter is displayed as 12", at which point the cumulative time and bandwidth decorrelation from 0.5 MHz channelization and 4s averaging reaches 10\%. }
\label{fig:pencilbeams}
\end{figure}

Thus, as long as low--resolution ``finder catalogues" are available, VLBI--resolution surveying is possible with DiFX2 with minimal overhead.  Figure~\ref{fig:pencilbeams} illustrates the use of a low--resolution image, and the VLBI datasets which would result from a multiple field correlator pass.  \citet{middelberg11a} have already used this new capability to carry out pilot VLBI survey observations in the Chandra Deep Field South, and these observations were instrumental in the development and refinement of the new correlation mode.  \S\ref{sec:performance} describes in detail the performance impact of adding multiple phase centers to a correlation.  \S\ref{sec:validation} shows the verification that $uv$ shifted visibility datasets have no residual phase or amplitude errors.  This feature is new in DiFX2.

\subsection{Correct model accountability}
\label{subsec:accountability}
The RPFITS output format used initially by DiFX1.x had no means to store an accurate representation of the delay model applied at the correlator.  The transition to FITS-IDI in version 1.5.0 has made correct model accountability possible, and an accurate representation of the applied delay model is now stored in two binary tables -- the ``IM" table and the ``MC" table.  These tables store the same sampled model polynomials used by DiFX2 and the applied clock model, and can be used by post--processing software such as AIPS\footnote{http://www.aips.nrao.edu/} to accurately make changes to the phase center of the correlated dataset, or correct for antenna position errors, earth orientation parameter errors and the like.

\subsection{New data monitoring tools}
\subsubsection{Autocorrelation filterbank ``spigot"}
\label{subsec:transients}
In order to facilitate searches for transient signals, an autocorrelation ``spigot" has been added to DiFX2.  This spigot supplies the autocorrelations from each antenna at user--specified time and frequency resolution, by means of a UDP multicast message.  The additional computational load is negligible, since the antenna autocorrelations are already calculated as a matter of course, and the additional load of sending the multicast messages is negligible for all but very short integrations.  The messages are sent with a simple plain text header, allowing (one or more) analysis programs to capture, time--order and inspect what are essentially N independent but time--aligned filterbank data streams, where N is the number of antennas.  This feature has been available since version 1.5.1, but is considerably improved in DiFX2.  An example of the two-dimensional dynamic spectrum which is obtained (for each antenna) is shown below in Figure~\ref{fig:acdump}.

At the VLBA, additional functionality has been added to allow feedback from the analysis programs, allowing them to request that small time ranges of baseband data be extracted after the correlation has finished and written to disk elsewhere.  Thus, the detection pipeline can trigger baseband data ``grabs" based on the autocorrelation filterbank data, permitting a detailed analysis at full time resolution after the correlation has completed.  For the first time, this offers the possibility of full--time, commensal observations on VLBI arrays. This facility is currently being used for a commensal transient search of VLBA data in support of the CRAFT fast transients project for the ASKAP telescope \citep{macquart10a}.  A full description of the VLBA fast transients pipeline will shortly be published by Wayth et al. (in preparation).

\begin{figure}
\begin{tabular}{c}
\includegraphics[width=0.98\textwidth]{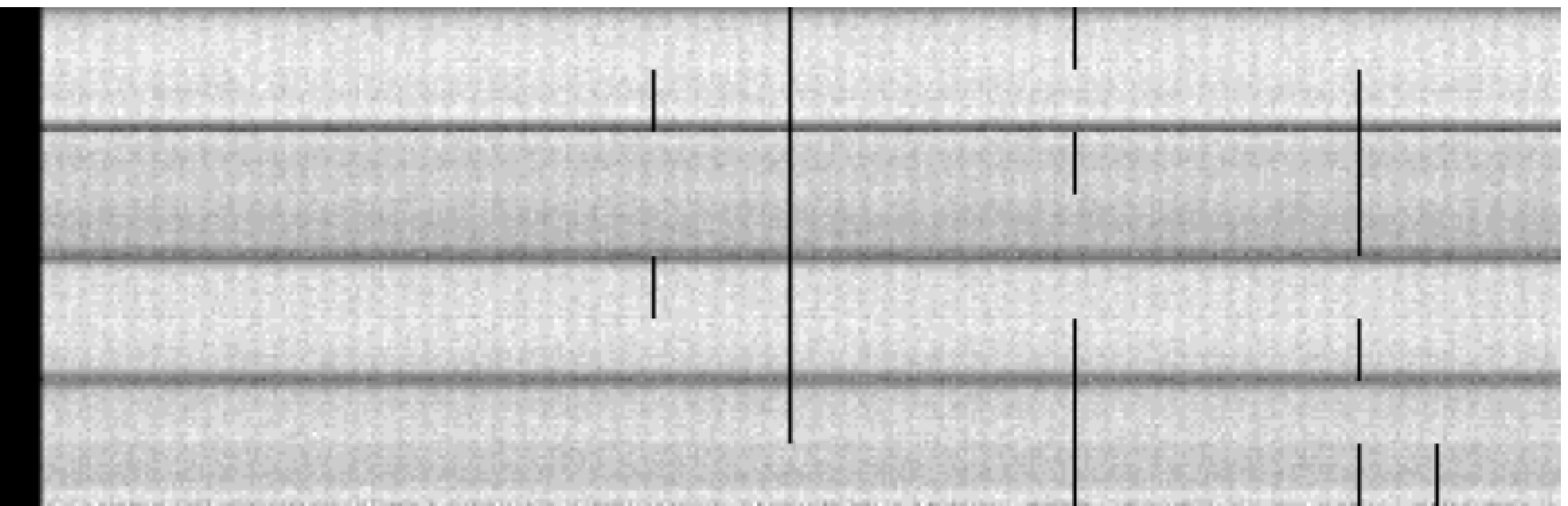} \\
\\
\includegraphics[width=0.98\textwidth]{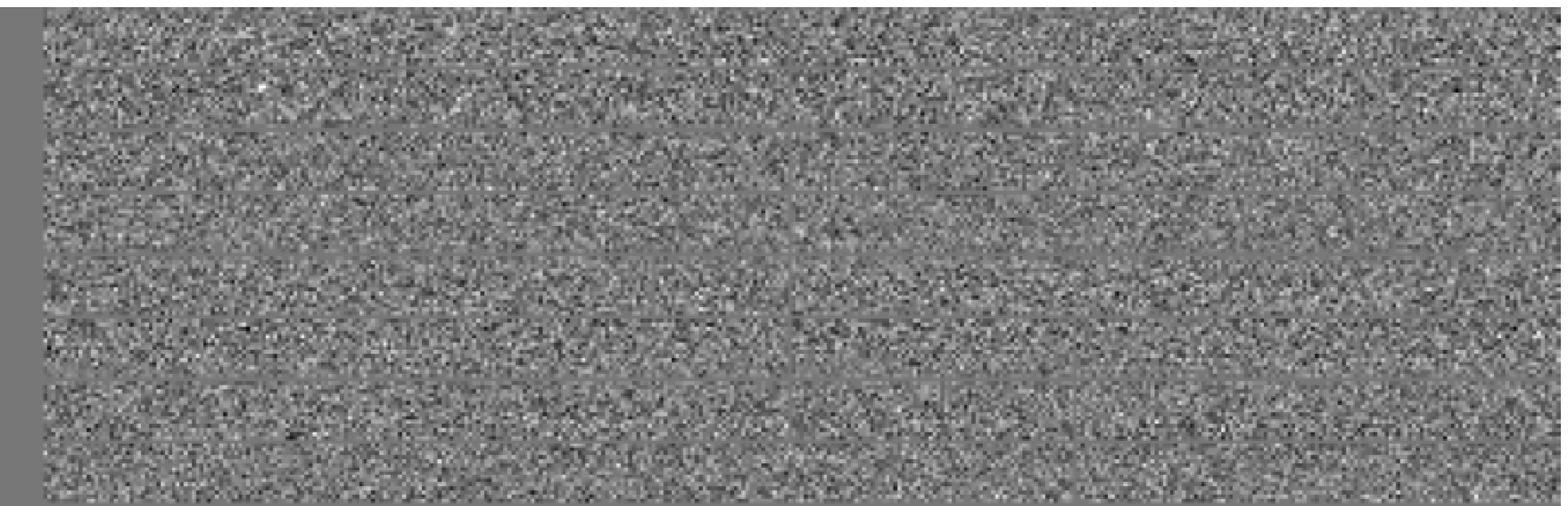}
\end{tabular}
\caption{Example of the autcorrelation dynamic spectrum produced from the Brewster VLBA antenna, with greyscale intensity representing autocorrelation signal strength.  These data were captured commensally during an observation in June 2010.  Time runs horizontally covering a period of 1 second, and the 64 MHz of bandwidth (consisting of eight concatenated 8 MHz sub--bands spanning 1350.49 -- 1414.49 MHz in right circular polarization) runs vertically.  A single pixel is 2 ms and 500 kHz.  {\bf (top)} The raw filterbank output.  The imprint of the 80 Hz noise calibration signal present in VLBA data is clearly visible, as is that of the sub--band bandpasses.  Data lost during times of network congestion appears as zero amplitude (vertical black lines). {\bf (bottom)} The processed filterbank data presented to transient detection code, which has been filtered to interpolate missing data, remove bandpass shapes and remove the noise calibration signal. }
\label{fig:acdump}
\end{figure}

\subsubsection{Station--based kurtosis estimation}
\label{subsec:kurtosis}
Spectral kurtosis \citep{nita07a} can be calculated for radio filterbank data to estimate the form of the probability density distribution function for each filterbank channel.  Since radio frequency interference (RFI) generally corrupts the form of the probability density distribution of the filterbank data, spectral kurtosis can be a powerful and cheap method of identifying spectrally confined RFI.  Recently, \citet{deller10a} tested a simplistic implementation of spectral kurtosis calculation within DiFX and used it to identify previously unnoticed, rapidly time varying RFI at one VLBA station.  The test implementation described in \citet{deller10a} has been updated to the fully correct spectral kurtosis calculation in DiFX2, and the results are now made available using the same ``spigot" architecture used for the autocorrelation filterbank described in \S\ref{subsec:transients} above.  The spectral kurtosis values are used to identify RFI before the filterbank data is passed through the transient detection pipeline.
%, and kurtosis plots are also made for the convenience of PI of the original proposal.}

\subsubsection{Real--time visibility monitoring}
\label{subsec:realtimevis}
The final new data monitoring tool included in DiFX2 is a TCP--based visibility monitor server, which was first deployed in version 1.5.2.  Enabling this features causes DiFX to send copies of the visibility data through a TCP network connection to a ``monitor server", which sorts and sends selected visibilities to connected clients for real--time processing and/or display.  This feature enables data quality assessment during correlation, which is particularly useful for the verification of correct array and correlator setup during non--disk based observations (``eVLBI").

\subsection{Other new functionality}
\label{subsec:other}
A number of other minor yet useful features have been added to DiFX2.  The first is the ability to compensate for local oscillator (LO) offsets on a frequency by frequency basis.  An LO offset at a station leads to continuously wrapping phase with time (constant across all spectral channels) on all baselines to the affected station.  In DiFX2 an appropriate phase compensation is made after channelization in the same operation as fractional sample correction \citep[see][]{deller07a}.  This allows correction of LO offsets up to a small fraction of the spectral channel bandwidth (so typically up to tens of kHz) without significant decorrelation occurring.  LO offset correction can be used in tandem with spectral selection to aid in correlating mismatched sub--bands, but at this time phase calibration tones cannot be extracted from sub--bands which have had an LO correction applied. The lifting of this restriction will be the subject of future development.

In addition to reading from files and Mark5 modules, DiFX has the ability to accept baseband data from a network socket.  In DiFX1.x, this data transfer over network was restricted to the use of a TCP (guaranteed transfer) transport protocol.  Whilst it was possible to obtain sufficiently good network performance over short and/or dedicated network links to perform high speed eVLBI with DiFX1.x \citep[see e.g.,][]{phillips07a}, the congestion control inherent in TCP makes it unsuitable for long, potentially lossy network transfers.  DiFX2 retains the ability to perform correlations reading from a TCP socket, but adds the ability to connect data sources via UDP - a transfer protocol without congestion control that is better suited to maintaining high transfer speeds on a long, lossy or shared transmission network.

DiFX2 now supports the reading of complex sampled baseband data (in the VDIF format only).   Complex sampled data (where the digital representation of the antenna voltage is stored in a complex representation at half the sampling rate of a real sample stream) offers potential advantages over real sampling, including less processing required before storage and slightly reduced quantization losses at a given bit precision.  Modern digital data acquisition systems typically use complex sampled data internally, but in all current VLBI systems (and many non--VLBI systems) the complex sample stream is converted to a real sample stream before recording/correlation.  Supporting complex sampled data will reduce the amount of work required for an ad--hoc experiment utilizing a non--VLBI antenna with a system which does not convert to real sampling, and prepares for potential future VLBI systems utilizing complex sampling.  As with real data streams, complex sampled data streams can be obtained from network connections, disks, or Mark5 modules, and can be correlated against other complex data streams or real data streams. 

Finally, DiFX2 has extended the flexibility of clock model specification.  Like most correlators, DiFX1.x possessed the ability to compensate for a clock offset and linear rate of change at each station.  DiFX2 allows the specification of an arbitrary order clock polynomial at each station, allowing more accurate correction for known clock variations.  This is likely to be most applicable at very high frequencies.

\section{Performance improvements}
\label{sec:performance}
Testing of the performance of DiFX has been carried out on a variety of Intel--based clusters. Most have been comprised of recent Intel Xeon multi--core CPUs -- the results presented here were derived from the 10 node cluster installed at the NRAO Domenici Science Operation Center in Socorro, NM, to replace the VLBA hardware correlator.  Each node in this system is dual--CPU, where each CPU is an Intel Xeon quad--core with 6 MB of shared L2 cache, running at 2.5 GHz.  Each node has 4 GB of RAM.  Including the 1 Gb ethernet switching infrastructure, the total cost of the cluster (in 2008) was approximately US\$30,000.  As shown below, this system is capable of sustaining throughput of 512 Mbps for 10 stations, twice that of VLBA hardware correlator for typical experiments.  As part of the ongoing VLBA sensitivity upgrade program\footnote{http://www.vlba.nrao.edu/memos/sensi/} (which aims to demonstrate a station data rate of 4 Gbps by 2011, along with routine operations at 2 Gbps) this cluster is being substantially enlarged, at a cost which is a small fraction of the cost of recording media required to operate at the higher data rates.  DiFX is well poised to take advantage of ongoing improvements in CPU technology, such as the trend to many--core architectures and extended instruction sets, through its use of IPP for vector operations.  IPP is regularly updated to make best usage of the latest Intel (and Intel--compatible) CPU architectures.

Two code changes in DiFX2 dominate the performance improvements over earlier incarnations of DiFX.  The first is a more efficient implementation of vector phase rotations where the phase change is constant from one element in the vector to the next.  In DIFX, such operations include:

\begin{itemize}
\item The (time domain) fringe rotation, which multiplies a constant oscillator frequency by a changing model delay \citep[see][]{deller07a}.  Over short time periods, the delay change from one sample to the next is linearly approximated, which yields a constant phase increment;
\item The (frequency domain) fractional sample correction, which multiples a constant (for a given time window) delay by the frequency of the spectral point \citep[see][]{deller07a}.  The linearly increasing frequency of the spectral points across the subband again yields a constant phase increment; and
\item The repositioning of phase centers in a multi--center correlation (DiFX2 only), which is conceptually identical to a fractional--sample delay, although the magnitude of the delay applied can be greater than one sample.
\end{itemize}

\begin{figure}
\includegraphics[width=0.9\textwidth]{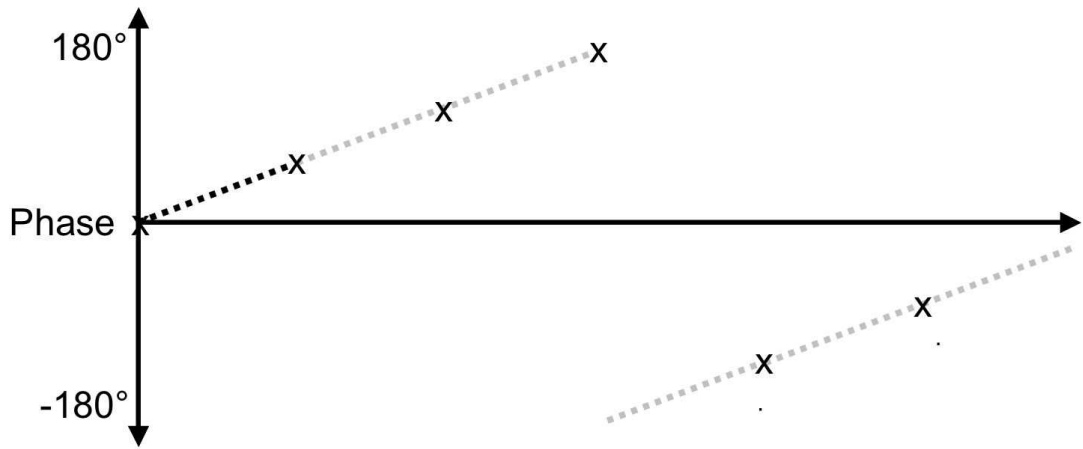}
\caption{Illustration of the complex multiplication ``filler" approach to minimise the trigonometric operations needed to produce a vector with linear phase change from one element to the next.  The figure plots the phase in degrees of a complex vector of unit amplitude -- the desired rotation vector.  The complex values shown by the black points and crosses are calculated trigonometrically, and the values shown by gray points are then filled in with a series of complex multiplications.  Each value shown by a cross is used in turn to rotate the vector of black points, and the resultant vector of complex numbers has an appropriate sequence of phases to be stored in the final rotation array between that cross and the next.}
\label{fig:stationbased}
\end{figure}

Phase rotations are applied to complex data using complex multiplications by a ``rotation" vector of unit amplitude, whose real and imaginary components are computed by taking a sine and cosine (sin/cos) of a vector containing the desired phase changes.  In general--purpose CPUs, the evaluation of trigonometric operations such as sin/cos is considerably more computationally expensive than complex arithmetic such as multiplication or addition.  When the phase interval change between array elements is constant, it is no longer necessary to compute sin/cos values for every array element, as was implemented in DiFX1.x.  Instead, as shown in Figure~\ref{fig:stationbased}, it is possible to compute the exact sin/cos values for only the first few elements of a vector (a ``subvector") and a series of offset elements.  The subvector is then multiplied by each offset in turn, and the rotated subvector is placed in an appropriate position in the final vector.  The number of trigonometric operations required could be further reduced by performing a similar decomposition of the subvector, but a cascaded approach such as this is not implemented to avoid numerical precision issues, and because the performance improvement from further decomposition is marginal for typical array sizes in DiFX.

With a suitable choice of the subvector length for this basic decomposition (typically the nearest factor of two to the square root of the vector length $V$) it is possible to reduce the required number of trigonometric computations by a factor of up to $\sqrt{V}/2$.  In the DiFX1.x implementation, trigonometric operations comprised almost a third of the station--based computational load.  In DiFX2, where fringe rotation and fractional sample vectors are typically of length 100--1000 elements, the resultant order--of--magnitude computational saving means that the cost of the trigonometric operations is reduced to a near--negligible value, at a small cost of extra complex arithmetic.  For correlations with a small number of stations ($\leq10$), the net increase in correlator throughput is 15--20\%.  (Alternatively, 15--20\% less computational resources are required to obtain a prescribed throughput).

The second efficiency improvement comes from an increase in flexibility in the order of traversing of the baseline--based cross--multiplications, allowing DiFX2 to mitigate the effect of cache misses when the size of the visibility result vector grows large (many stations and/or many spectral channels).  A cache miss occurs when data is no longer available for immediate use in the CPU cache (after being overwritten by some more recently used data), and must instead be retrieved from main memory, which is a relatively slow operation.  In DiFX1.x, every polarization pair of every frequency sub--band for every baseline at a given time was cross--multiplied sequentially.  When the visibility result array grows too large to remain in cache, this traverse leads to a cache miss for every accumulation, and a dramatic slowdown.

In DiFX2, the results of $N$ channelizations (hereafter referred to as FFTs [Fast Fourier Transforms] in accordance with the implementation) are buffered for each datastream, and a single selection of spectral points for a given baseline are then cross--multiplied $N$ times into the same area of the visibility result buffer.  This allows the intermediate vectors at the datastream to remain in cache during the station--based processing for $N-1$ passes, and the visibility result vector to remain in cache for $N-1$ passes.  For moderate values of $N\sim10$, a minimal amount of extra memory is required and the number of cache misses is greatly reduced.

\begin{figure}
\includegraphics[width=0.9\textwidth]{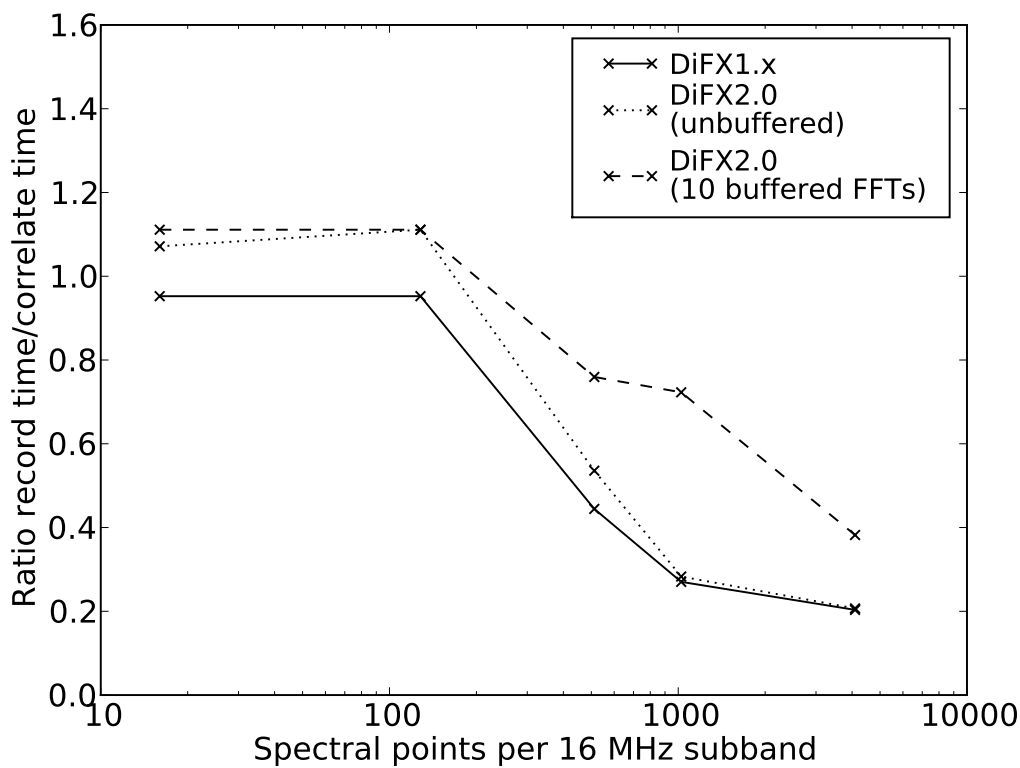}
\caption{Throughput of DiFX2 (with and without FFT buffering) compared to DiFX1.x for a varying number of spectral points (quoted per 16 MHz sub--band).  The test was run on the VLBA DiFX cluster (10 dual quad--core nodes) on 9 stations of data totaling 512 Mbps/station (4 frequencies, dual polarisation = 8 sub--bands).  Four polarisation products were computed for each of the frequency bands.  The vertical axis shows the ratio of record time to correlate time (the speed-up factor of the correlation).  For values above 1.0, the correlation is proceeding faster than the data was originally recorded.  The DiFX1.x performance (solid line) drops sharply when the visibility results exceed the available node cache, as does DiFX2 with no FFT buffering (dotted line), but DiFX2 with FFT buffering (10 FFTs buffered; dashed line) is much less adversely affected.}
\label{fig:baselinebased}
\end{figure}

Figure~\ref{fig:baselinebased} shows the net improvement in throughput for DiFX2 compared to DiFX1.x, for varying numbers of spectral points (and hence varying visibility result lengths).  The improvement is less marked for small numbers of spectral points, reflecting the 15--20\% improvement solely from the more efficient trigonometric processing.  As the visibility result buffer exceeds the cache size (each processing thread had access to approximately 1.7 MB of L2 cache, which the visibility buffer exceeds in size for 512 spectral points) the older DiFX1.x code suffers a marked drop in performance.  DiFX2 also experiences reduced throughput (primarily due to the increased FFT cost with a larger number of spectral points) but the reduction is much smaller.

\begin{figure}
\includegraphics[width=0.9\textwidth]{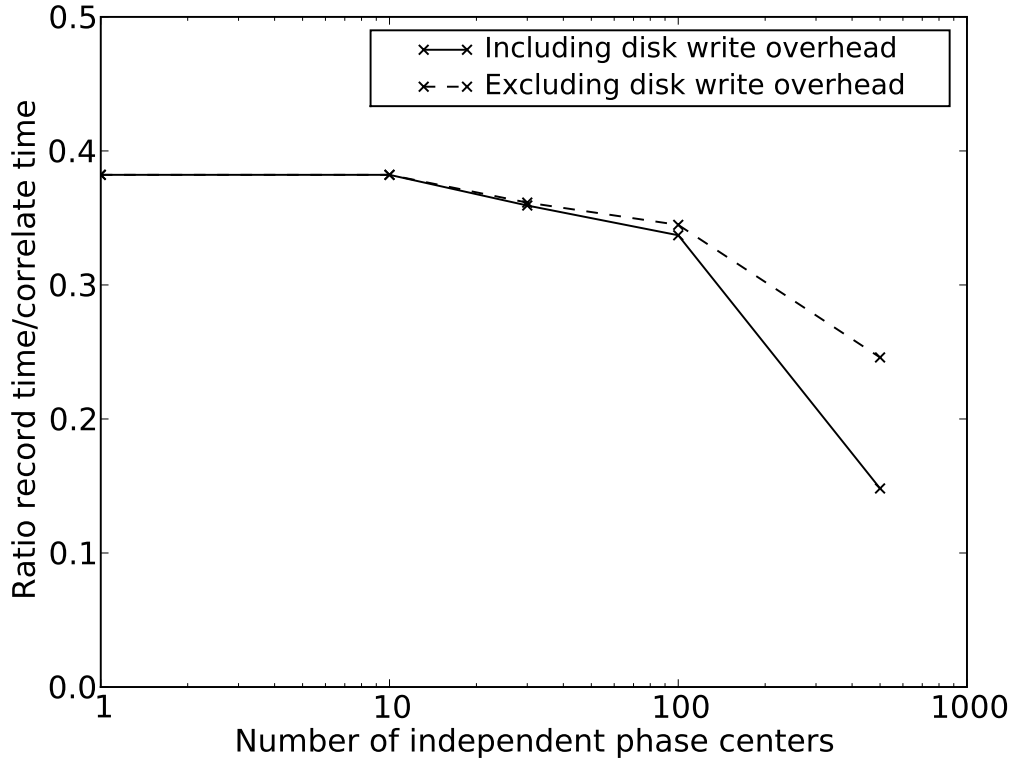}
\caption{Throughput of DiFX2 (10 buffered FFTs) for an increasingly large number of phase centers.  The dataset and correlator resources were identical to the previous benchmark, and as before, the vertical axis shows the ratio of record time to correlate time.  4096 spectral channels per band were used, and the $uv$ shifts were performed every 26 ms.  All polarisation products were computed. The solid line shows the observed throughput, while the dashed line removes the slowdown caused by writing the visibilities to a slow disk.  Up to hundreds of phase centers can be correlated whilst imposing a near--negligible impact on correlator throughput, although disk write speed becomes a limitation with hundreds of fields. This could be mitigated with a faster RAID disk for storing output visibilities.  Even combined with the slowdown due to the larger FFT size (as seen in Figure~\ref{fig:baselinebased}), the correlation cost of producing five hundred phase centers is only 4.5 times that of a normal, single phase center continuum correlation -- yielding a speed-up factor greater than 100.}
\label{fig:mpcbench}
\end{figure}

Finally, the performance of DiFX2 with multiple phase centers should be noted.  Figure~\ref{fig:baselinebased} shows that the high spectral resolution required to minimize the decorrelation suffered during $uv$ shifts carries its own penalty (dependent on the sub--band bandwidth, but typically 2048 or 4096 spectral points, yielding a computational load increase of 2--3x over a standard 16 spectral point continuum observation).  However, beyond this initial penalty, the cost of adding additional phase centers is very small.  Figure~\ref{fig:mpcbench} shows the variation of correlator throughput with number of phase centers for a fixed spectral resolution. For this test, spectral resolution of 2 kHz and temporal resolution of 26 ms was used -- sufficient to shift to the edge of the VLBA primary beam at 1.4 GHz (15') with $<$5\% decorrelation due to time and bandwidth smearing.

At very large numbers of phase centers, the visibility data rate to disk becomes large, even with heavy spectral and temporal averaging.  In this test, 16 spectral points per band and 4 second averaging was used, yielding a data rate per phase center of 70 kB per second of recorded data.  For 500 phase centers, this makes a total output data rate of 35 MB per second of recorded data, which is not inconsiderable when the overhead of writing to an array of different output files is considered.  This limitation can be overcome with minimal outlay by writing the output visibilities to commercially available high speed, low latency disk arrays, and by using a file system optimized for large numbers of file operations.

\section{Operational infrastructure}
\label{sec:operations}
Considerable effort has been expended to improve the usability of DiFX in routine operations.  Specifically, the configuration of correlator jobs has been simplified, as has the generation of the correlator model, and extensive monitoring and logging has been added.  Some elements of the new infrastructure are specific to the VLBA installation, but can be easily customized in many cases to suit the needs of a different installation.  All of the packages described below were developed in the latter stages of the DiFX1.x series and are available both in DiFX-1.5.4 and DiFX2.

\subsection{Correlation configuration}
\label{sec:vex2difx}
The pathway for automatic configuration of the correlator control files has been considerably improved since DiFX1.x.  A new program ``vex2difx" can populate the entire set of necessary correlator files based only on the ``vex"\footnote{http://www.vlbi.org/vex/} observation description file, while default values for parameters such as integration time and spectral resolution can be overridden as desired.  For the VLBA installation of DiFX2, ancillary information not available at scheduling time such as data module names, Earth Orientation Parameters, stations clocks, etc is provided automatically to vex2difx, making use of the operational database available at the VLBA.  For other arrays, these necessary inputs can be provided to vex2difx by hand or using a similarly customized script.

\subsection{Model generation}
The model generator used by DiFX1.x, which was based on a customized implementation of CALC\footnote{http://gemini.gsfc.nasa.gov/solve/} with limited support, has been surpassed by a more flexible client/server architecture.  In this approach, the DiFX model file writing is handled by a stand--alone program which is completely divorced from the CALC--based delay calculations, which are performed with a standard installation of CALC9 and communicated upon request.  This ``calcserver" was already widely used at VLBI observatories including the VLBA and MPIfR, and has now been adopted by the LBA.  Accordingly, it has the advantage of easier integration with existing observatory setups.  Concurrently with this change, DiFX2 has been enhanced to directly read the polynomial--based delay models generated by the model client and stored in the FITS IM and MC tables.  This ensures a perfect match between the recorded and applied geometric model.  In contrast, DiFX1.x read sampled delay files and used a low order interpolation between the sampled points, which resulted in errors of the order of a tenth of a femtosecond.  These very small errors were discovered in the detailed comparisons presented in \S\ref{sec:validation}.  In addition, the use of a polynomial--based model representation saves disk space and memory, as it is more compact than the sampled--delay representation used by DiFX1.x.

\subsection{Correlation monitor, control and archiving}
As part of the DiFX2 development effort, a standardized message package (``difxmessage") was created.  This uses multicast xml messages to broadcast the state of various resources involved with a correlation, as well of the progress of the correlation itself.  Resource messages include the CPU and network utilization, and potentially error messages describing equipment failure or misconfiguration.  The messages are graduated in importance from ``debug" to ``fatal".  A configurable display and logging program for these messages has been created to meet the inspection needs of different users.

This message package is also used by other programs available in DiFX2 to enable an immediate halt to correlation, the quarantining of resources, and other useful miscellaneous tasks.  VLBA operations has developed a number of tools specific to the VLBA installation, including a graphical user interface which allows correlator jobs to be queued and monitored.  These site-specific utility programs are also available as a starting point for adaptation to local needs.

\section{Validation testing}
\label{sec:validation}
The initial release of DiFX was tested against three correlators in a selection of observing modes \citep{deller07a,tingay09a}.  During its adoption by NRAO, DiFX1.x was subjected to much more extensive testing and comparison against the VLBA hardware correlator.   As with the earlier tests, the NRAO validation scheme for DiFX was primarily composed of point-by-point visibility comparisons, but many more recording modes (combinations of bandwidths, spectral resolution, and integration time) were compared, and unlike earlier validation correlations also included a number of ``functional" tests, where the final observable from astrometric or geodetic observations were compared between correlators. The VLBA DiFX test plan is described in detail by \citet{romney09a}.  

\begin{figure}
\begin{tabular}{cc}
\includegraphics[width=0.49\textwidth]{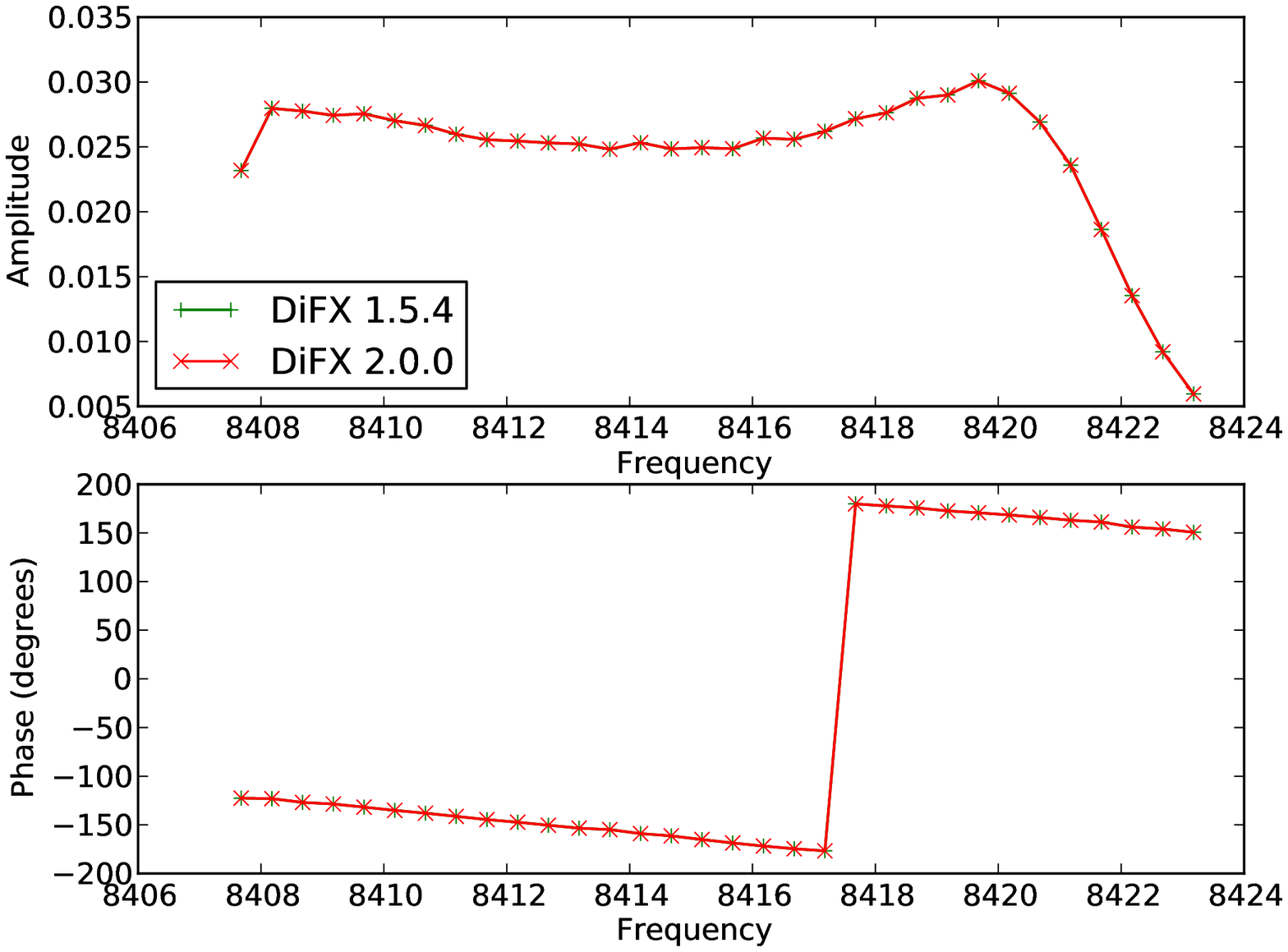} &
\includegraphics[width=0.49\textwidth]{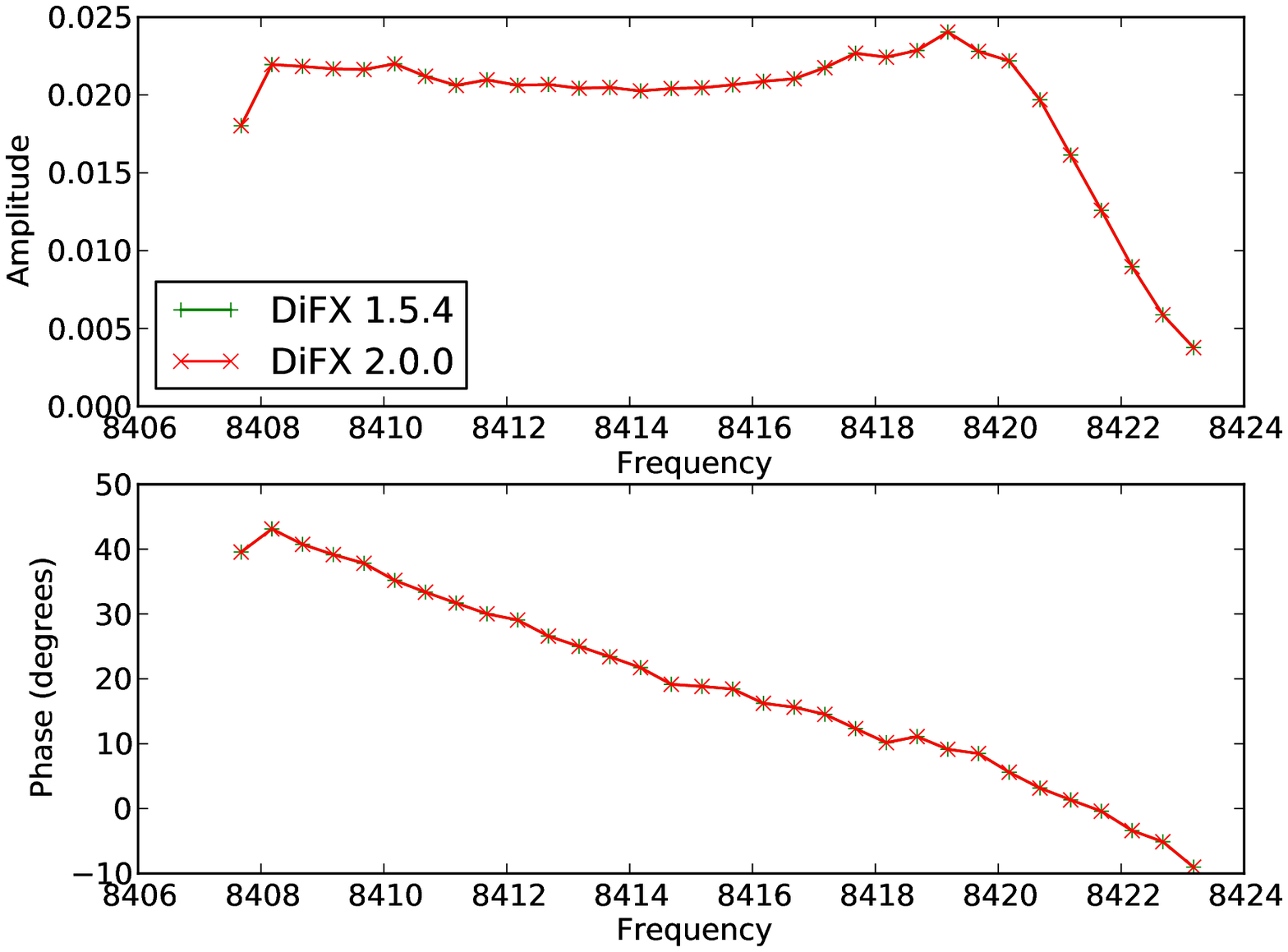}  \\
\includegraphics[width=0.49\textwidth]{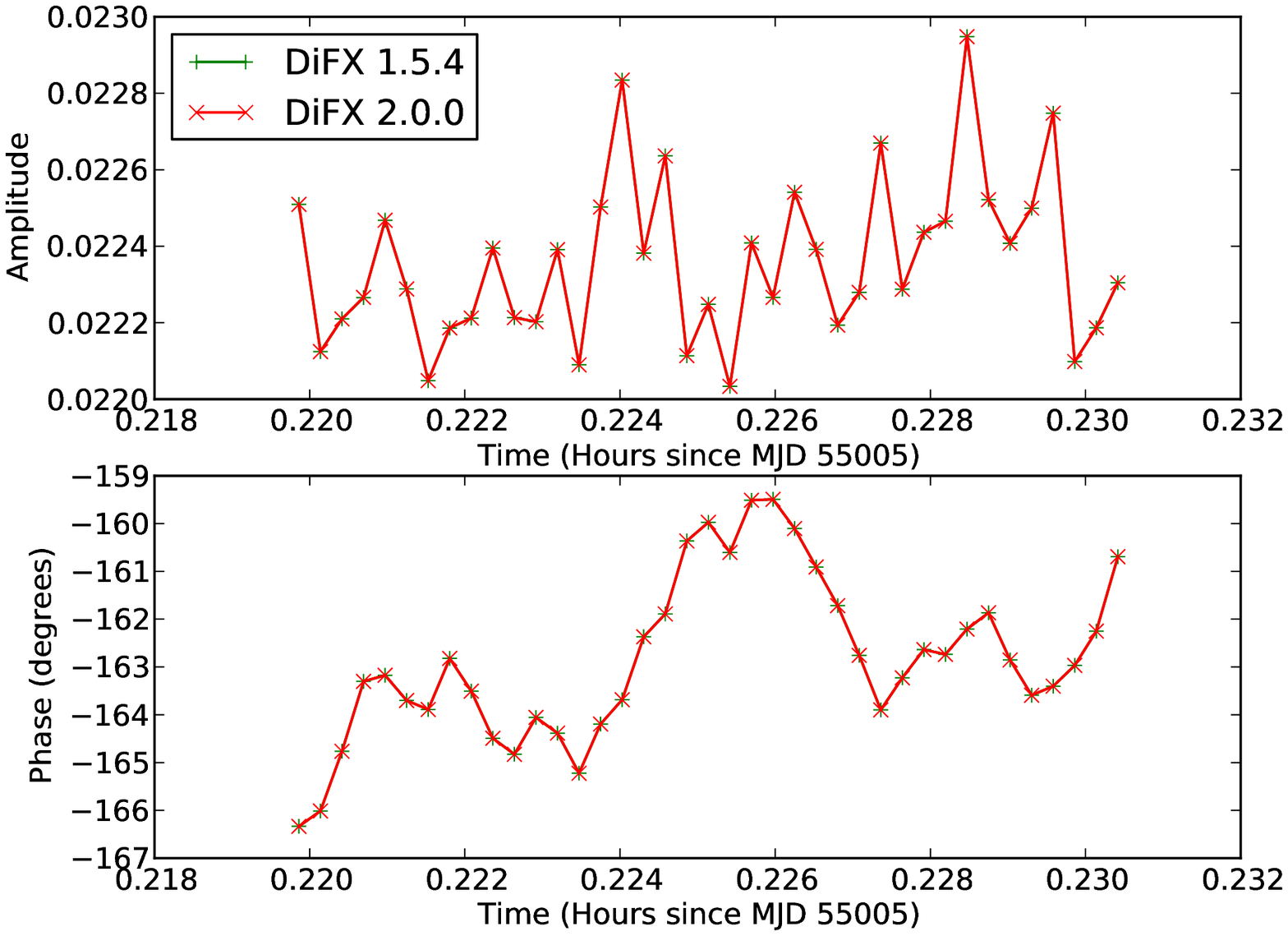} &
\includegraphics[width=0.49\textwidth]{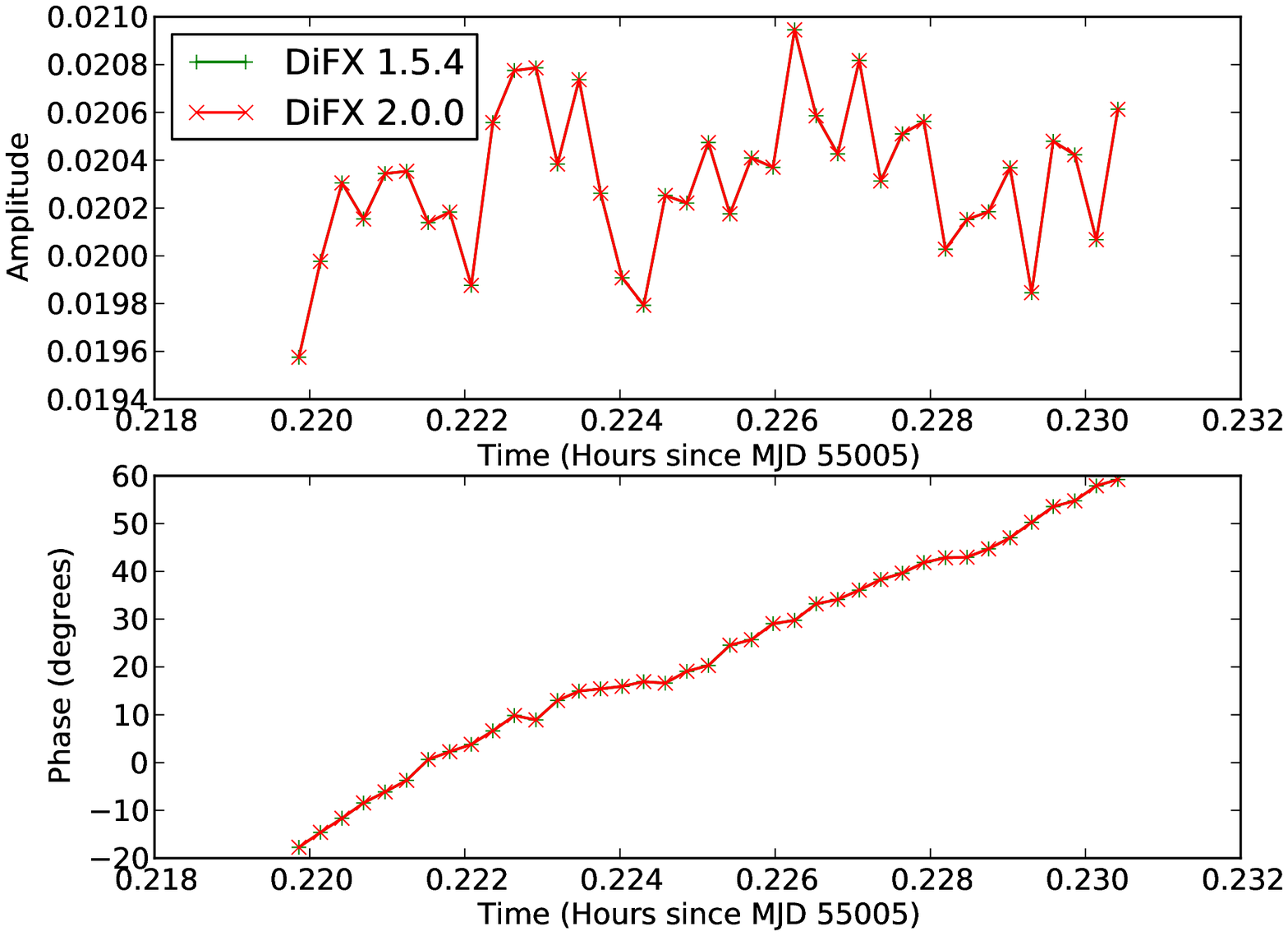} 
\end{tabular}
\caption{Comparison of the correlated output of DiFX2.0.0 (solid line) and DiFX1.5.4 (dashed line) for a single observing band for a pair of VLBA baselines over a 40 second scan. The target is the strong calibrator 4C39.25, and the observing band plotted spans 16 MHz from 8407.49 -- 8423.49 MHz, in right circular polarization.  The top row shows the baseline from Brewster to Fort Davis (2350 km) and the bottom row shows Brewster to Saint Croix (5770 km).  {\bf (left)} Visibility amplitude and phase (averaged across the band) versus time.  {\bf (right)} Visibility amplitude and phase (averaged for the scan duration) versus frequency.   No difference is apparent on this scale - formal analysis shows that the rms amplitude difference is 0.0007\% and the rms phase difference is 0.0007\degrees.}
\label{fig:comparison1}
\end{figure}

By the time of the adoption of DiFX2, the VLBA hardware correlator had already been retired and so the primary validation of the DiFX2 correlator was undertaken against the operating DiFX1.x installation at NRAO.  At the time of the tests, the specific versions in use were DiFX 1.5.4 (production) and DiFX 2.0.0 (testing).  A representative series of comparison plots is shown in Figure~\ref{fig:comparison1}, detailing the excellent agreement between the correlators.  The comparison shown here utilized 40 seconds of data on the bright calibrator 4C39.25 (J0927+3902), with one 16 MHz wide recorded band spanning the frequency range 8407.49 -- 8423.49 MHz.  The integration time was one second.  A 256 point FFT was used, with the resultant 128 spectral points averaged down to 32 in the FITS file.  These 32 spectral points were further averaged across the whole band for a statistical analysis, which showed that the rms phase deviation between the two correlators was 0.0007\degrees, and the rms amplitude deviation was 0.0007\%.

\begin{figure}
\begin{tabular}{cc}
\includegraphics[width=0.49\textwidth]{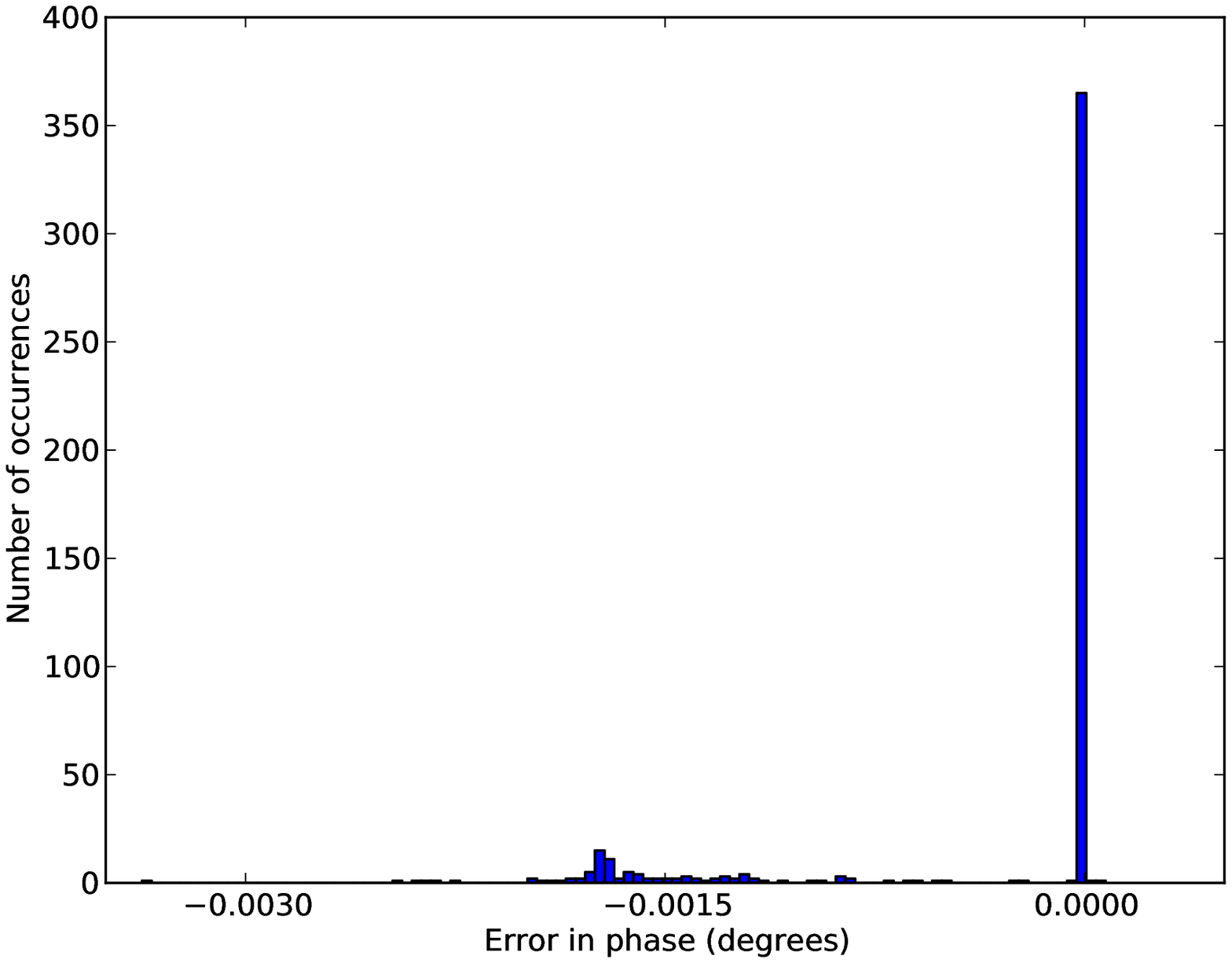} &
\includegraphics[width=0.49\textwidth]{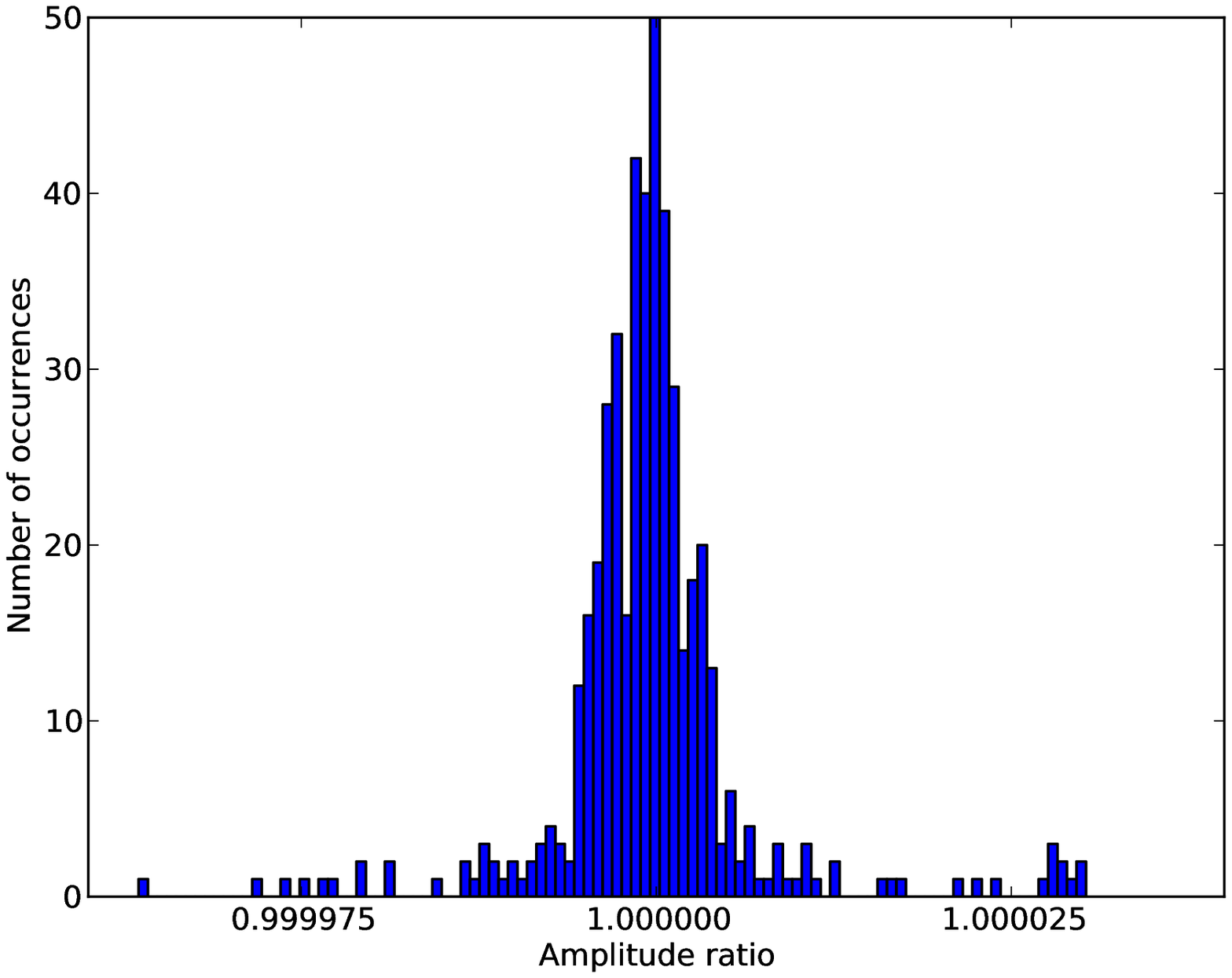}  
\end{tabular}
\caption{Comparison of the correlated output of DiFX 2.0.0 and DiFX 1.5.4 for a single observing band for a pair of VLBA baselines over a 40 second scan. The same baseband data was used as in Figure~\ref{fig:comparison1}.  {\bf (left)} Histogram showing phase difference between the two correlator outputs.  0.002 degrees corresponds to a delay error of approximately 10$^{-16}$ seconds.  {\bf (right)} Histogram showing fractional amplitude difference between the two correlator outputs.}
\label{fig:comparison2}
\end{figure}

Histograms of the errors in phase and fractional amplitude are shown in Figure~\ref{fig:comparison2}.  The histogram shows that the phase errors, whilst exceedingly small, do not have zero mean in this instance (the mean amplitude error, on the other hand, is less than a tenth of the rms).  This discrepancy is due to delay errors at the 0.1 femtosecond level in DiFX1.x, which used a 2nd order delay interpolator at 1 second timescales (DiFX2 uses a 5th order interpolator at 2 minute timescales, which is considerably more accurate).  It is worth noting that errors of this magnitude cannot even be discerned in comparisons between DiFX and hardware correlators such as that formerly used by the VLBA, due to the coarser fringe rotation and internal precision used by the hardware correlator.

\begin{figure}
\begin{tabular}{cc}
\includegraphics[width=0.49\textwidth]{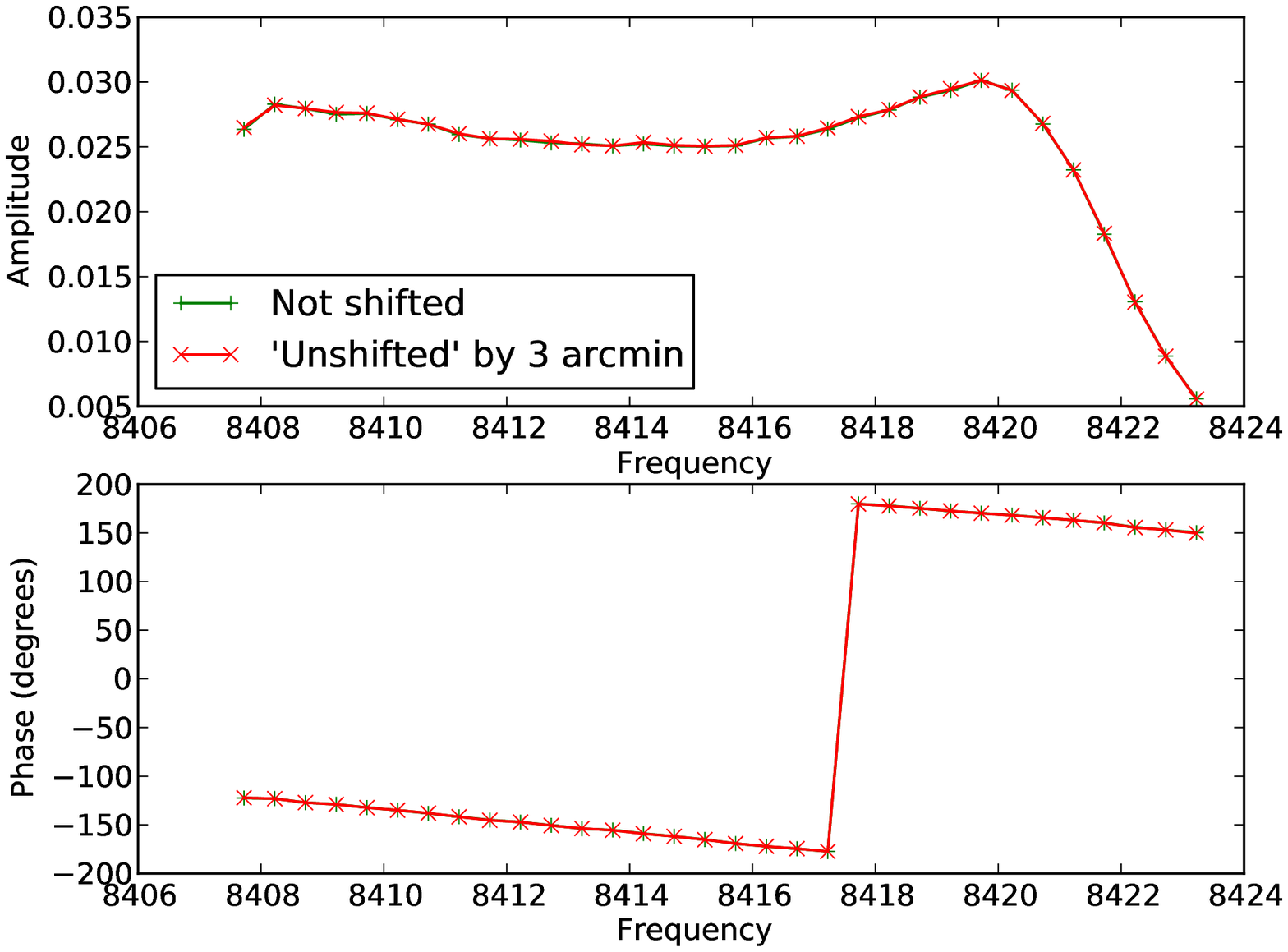} &
\includegraphics[width=0.49\textwidth]{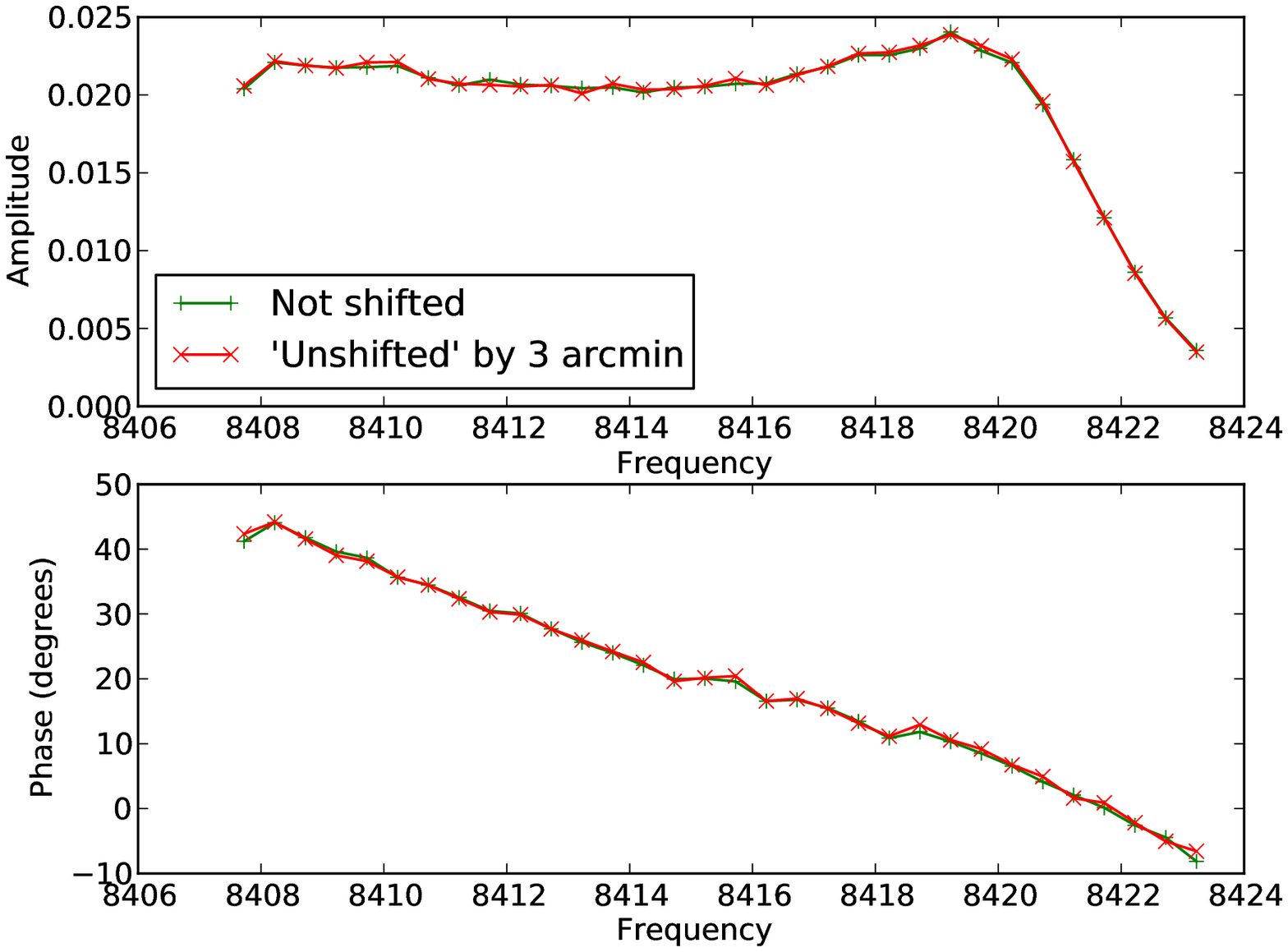}  \\
\includegraphics[width=0.49\textwidth]{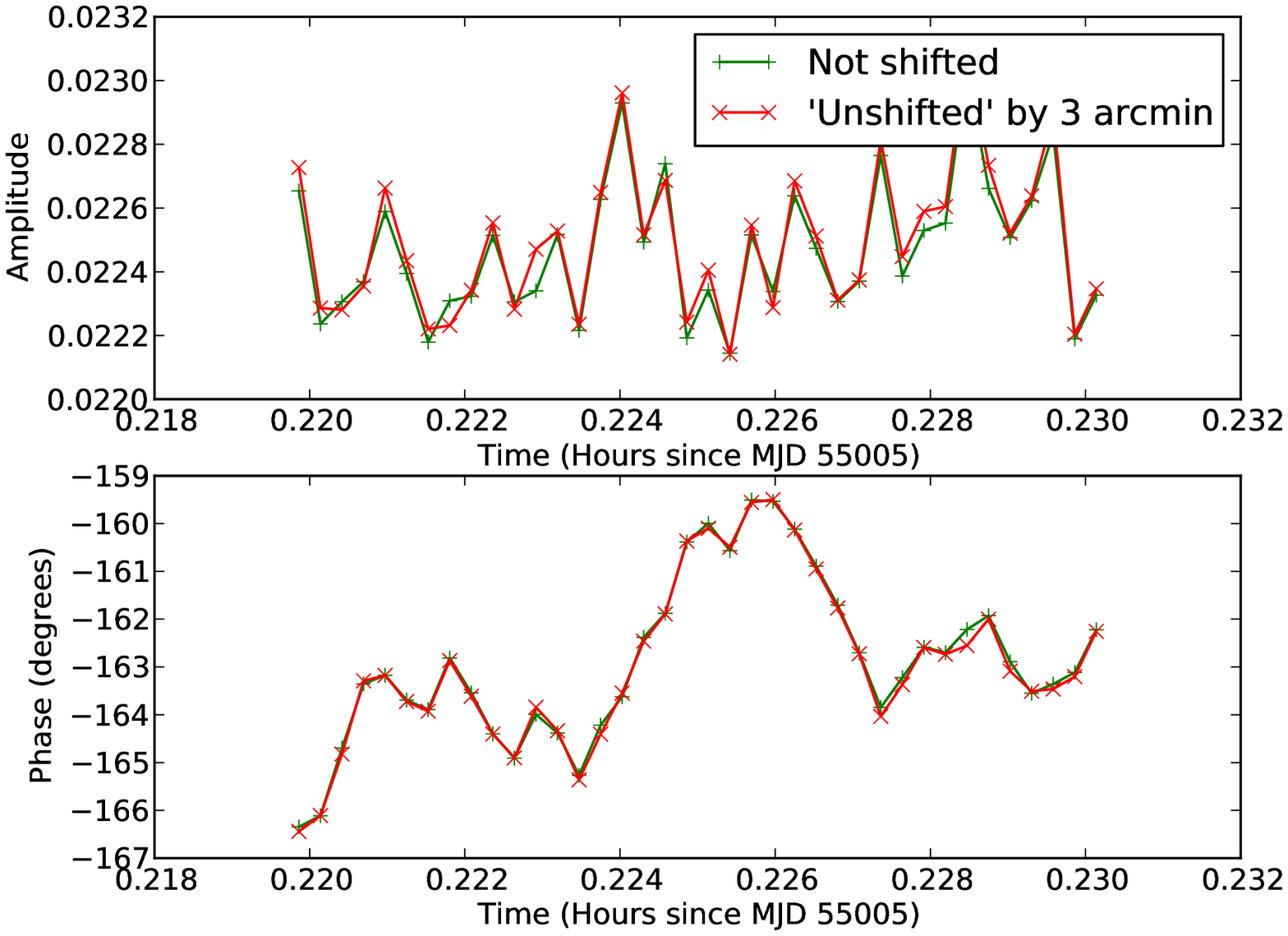} &
\includegraphics[width=0.49\textwidth]{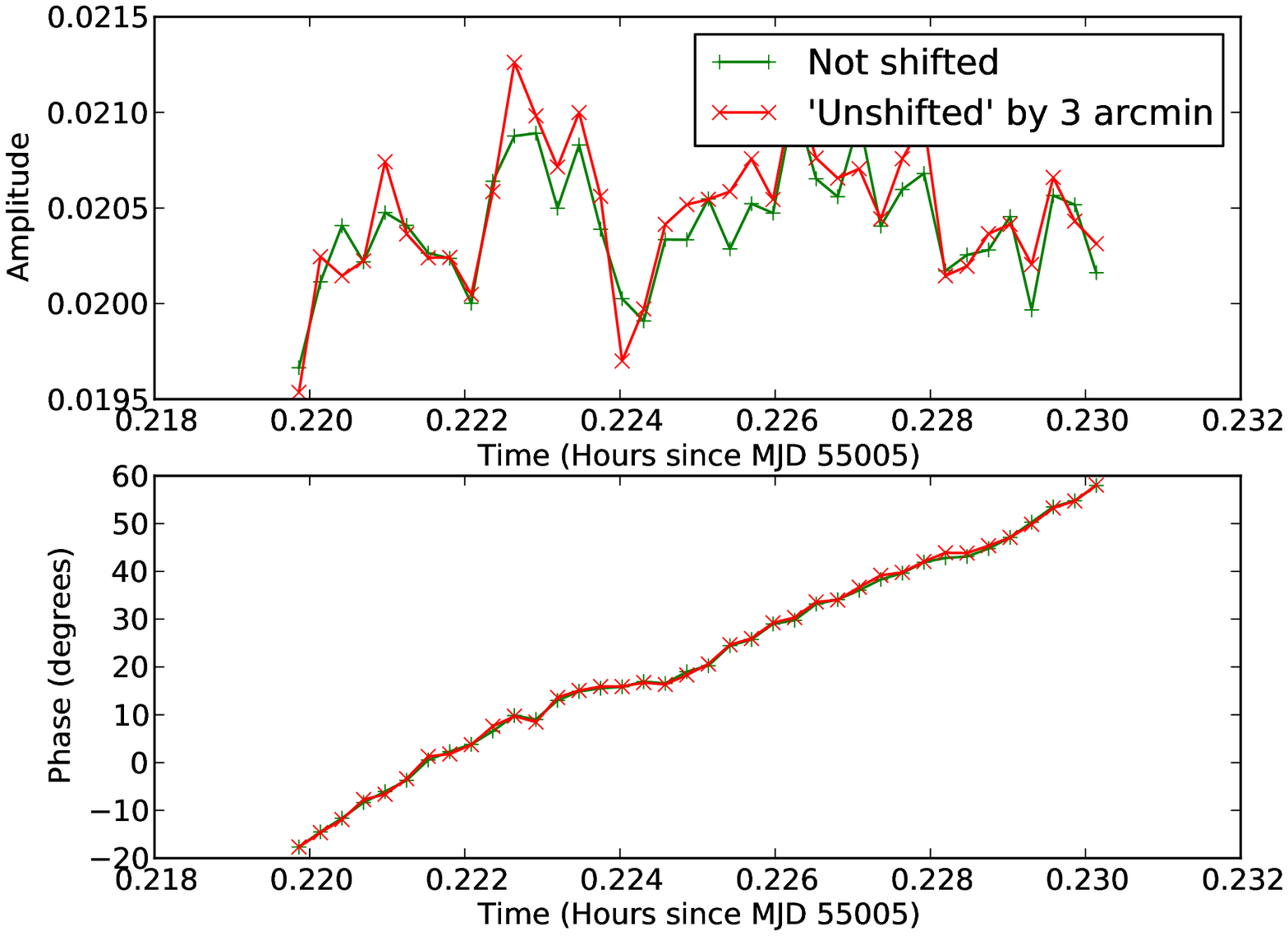} 
\end{tabular}
\caption{Comparison of the correlated output of an ordinary correlation (solid line) and one where the visibilities have been shifted back from an initial correlation center 3 arcminutes away (dashed line) for a single observing band for a pair of VLBA baselines.  The same baseband data and baselines were used as for Figure~\ref{fig:comparison1}.    {\bf (left)} Visibility amplitude and phase (averaged across the band) versus time.  {\bf (right)} Visibility amplitude and phase (averaged for the scan duration) versus frequency.  The rms amplitude difference is 0.09\%, and the rms phase difference is 0.014\%.  For each baseline the sensitivity loss can be roughly estimated using the ratio of the rms of the visibility amplitude over time.  For the Brewster to Fort Davis baseline the sensitivity loss is $\sim$3\% (against an expectation of 2\%), while for Brewster to Saint Croix the sensitivity loss is $\sim$25\% (expectation 23\%).}
%% The directory with these files on marathon is /home/adeller/testing/2.0/4c39[-shift]
%% The prefixes are comparison.1chan/comparison.32chan, and comparison.reshift.1chan/comparison.reshift.32chan
%% See the readme file in this directory
\label{fig:shiftcomparison1}
\end{figure}

\begin{figure}
\begin{tabular}{cc}
\includegraphics[width=0.49\textwidth]{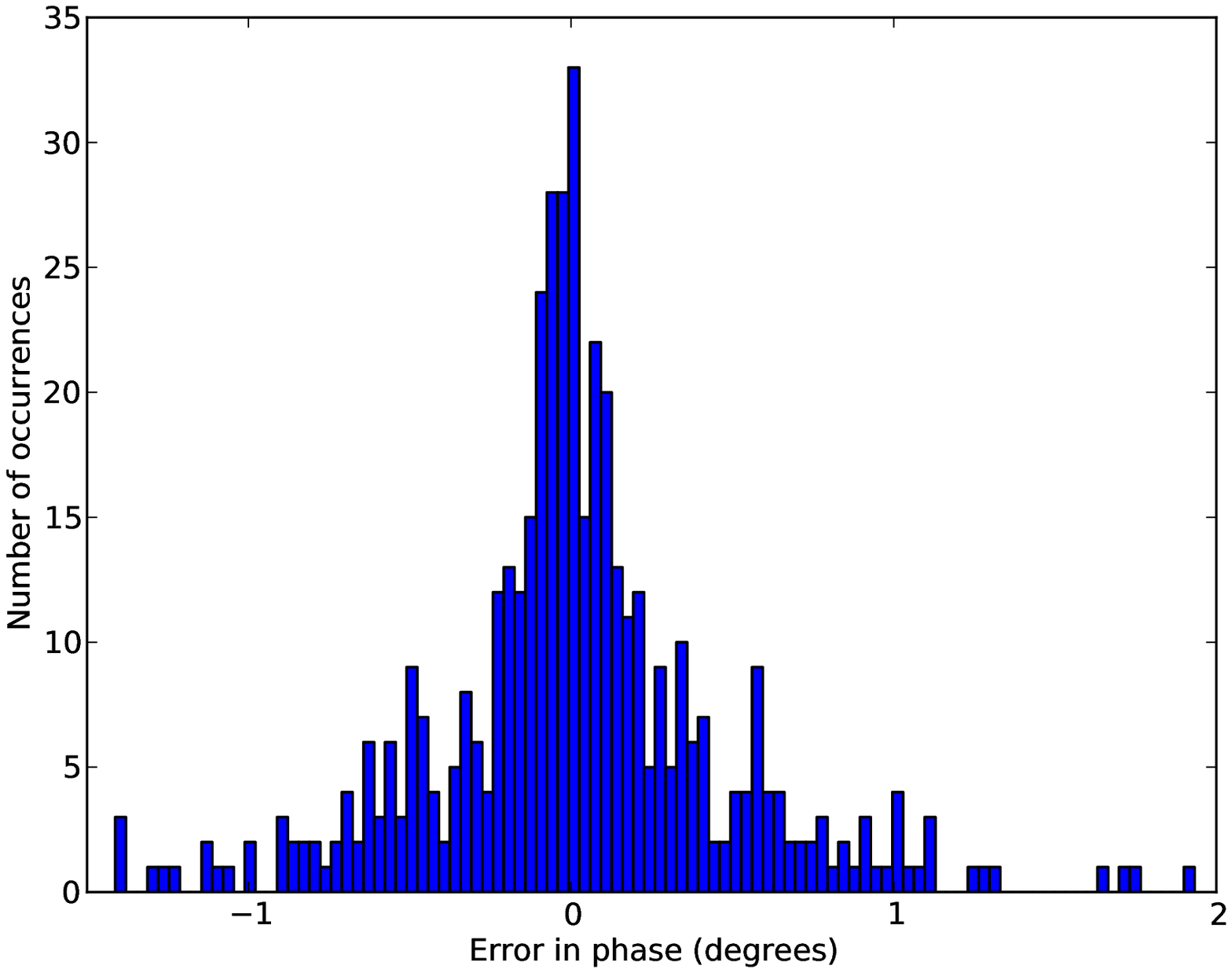} &
\includegraphics[width=0.49\textwidth]{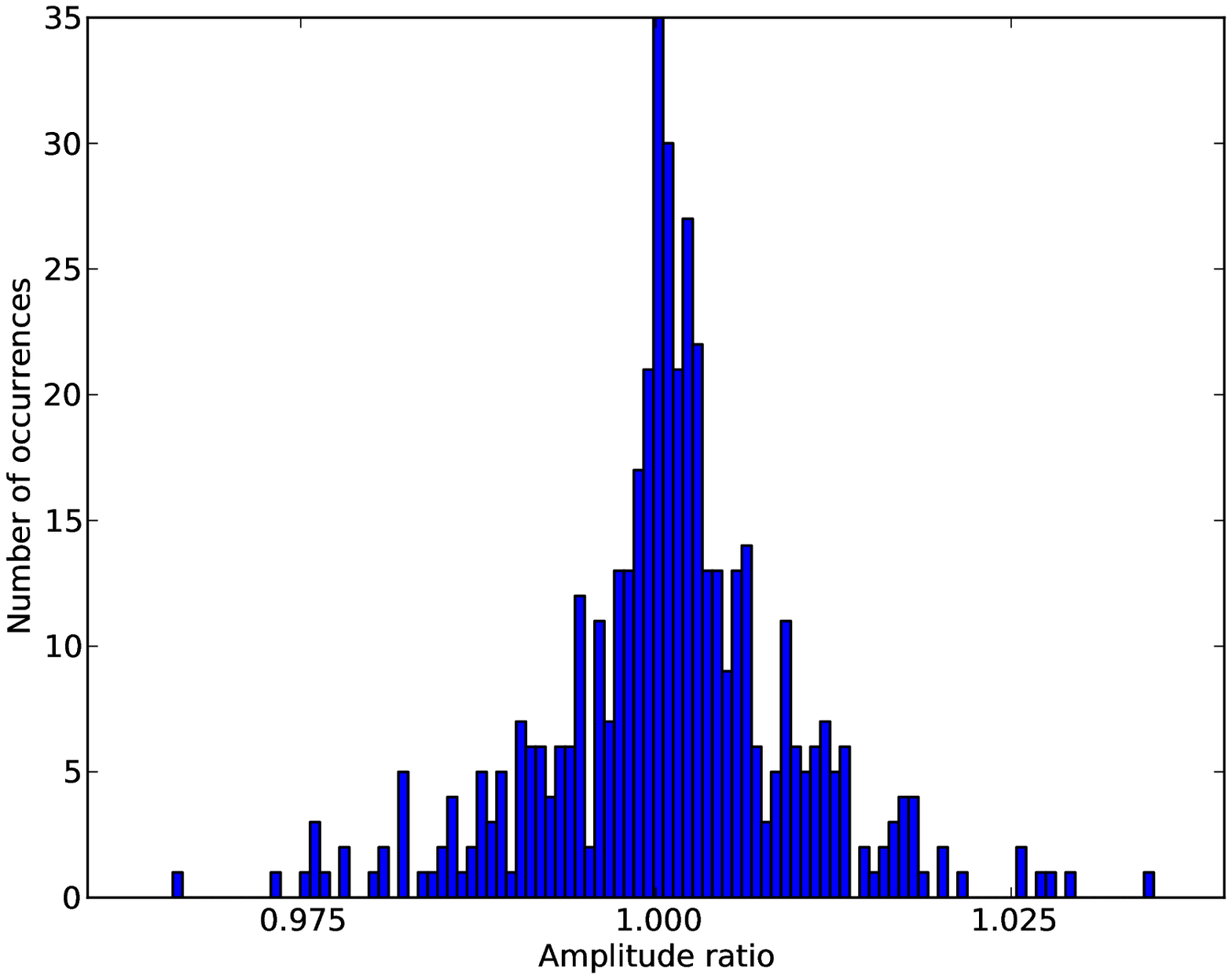}  
\end{tabular}
\caption{Comparison of the correlated output of an ordinary correlation and one where the visibilities have been shifted back from an initial correlation center 3 arcminutes away for a single observing band for a pair of VLBA baselines.  The same baseband data and baselines were used as for Figure~\ref{fig:comparison2}.  {\bf (left)} Histogram showing phase difference between the two correlator outputs.  {\bf (right)} Histogram showing fractional amplitude difference between the two correlator outputs.}
\label{fig:shiftcomparison2}
\end{figure}

This same 40 second time range of data was also used to verify the correct functioning of the multiple phase center code.  The correlation center for 4C39.25 was shifted by 2 arcminutes in declination and 2 arcminutes in right ascension, and a phase center was added at the true position of 4C39.25.  The FFT size was increased by a factor of four to 1024, and the $uv$ shifts were applied at a maximum interval of 40 ms.  However, the final spectral and temporal resolution were left unchanged at 0.5 MHz and 1 second, respectively.  A $\sim$3 arcminute shift leads to differential baseline delays of up to $\sim$10 microseconds for the VLBA.  For the correlation parameters chosen, this results in time and bandwidth decorrelation of 30\% and 7\% respectively.  Obtaining higher ``upfront" time and frequency resolution to greatly reduce the smearing would be straightforward, but these modest parameters were chosen to illustrate the correctness of the phase and amplitude compensation.  Figures~\ref{fig:shiftcomparison1} and \ref{fig:shiftcomparison2} below repeat the visibility--to--visibility comparisons previously made for the DiFX 1.5.4 to DiFX 2.0.0 comparison.

Across all six baselines used for the test, the mean amplitude and phase error was 0.09\% and 0.014\degrees\ respectively.  In each case, this is less than a tenth of the rms deviation observed in these quantities (0.9\%, 0.46\degrees).  Taking the rms of the visibility amplitude over time (per baseline) as a proxy for sensitivity, the decorrelation due to time and bandwidth effects can be estimated.  For each baseline, the decorrelation estimated in this manner is consistent with the predicted values to within several percent, and the average across all baselines agrees to 0.4\%.  Taken in conjunction with the perfect amplitude agreement (after the application of amplitude corrections for decorrelation), this shows that the $uv$ shift is performing exactly as expected.

\section{Future work}
\label{sec:futurework}
The development of DiFX2 will continue past the 2.0.0 version described in this paper.  Planned new functionality includes the ability to form one or more phased array outputs, as an alternative (or in addition) to a normal cross--correlation.  This would allow DiFX2 to produce tied--array beams for high time-resolution studies, such as pulsar analysis.  Both a digital filterbank (with tunable integration length) and a reconstructed time series are planned to be selectable outputs from the phased array.  A VLBI--capable phased array system is necessary for the Large European Array for Pulsars (LEAP)\footnote{http://epta.jb.man.ac.uk/leap.html} project, and DiFX may be used for this application.

Other new functionality under investigation includes frequency--division multiplexing for improved performance with larger numbers of antennas, efficient support for numbers of spectral points per sub--band which are not a power of two, and expanded graphical correlation monitoring and display.  As has been the case to date, development will be driven by the needs of the DiFX community, and it is likely that future applications will arise which are not presently envisaged.

\section{Conclusions}
\label{sec:conclusions}

A number of significant improvements have been made to the DiFX software correlator since its public release in 2007.  These have encompassed improved robustness, greater performance, and the addition of several valuable new features.  In particular, the DiFX2 series now supports phase calibration tone extraction, multiple simultaneous phase center correlation, and the production of high--time resolution filterbank and kurtosis data for use in transient searches and RFI mitigation.  In certain areas of parameter space, such as deep VLBI surveys, these new features allow processing speed improvements in excess of a factor of 100.  Collectively, all of these new features reinforce the advantages that software correlators possess over custom--designed hardware correlators.  DiFX2 has been adopted by three major VLBI correlator facilities for production usage, and numerous other institutes and individuals for experimental use. DiFX was a key element facilitating major bandwidth expansions at both the LBA and VLBA, and offers the chance for expanded resource sharing and improved robustness in worldwide VLBI.  Development of DiFX is expected to continue in the future, with the possibility of application to more existing and upcoming arrays.

\acknowledgements  ATD is supported by an NRAO Jansky Fellowship.  The International Centre for Radio Astronomy Research is a Joint Venture between Curtin University of Technology and The University of Western Australia, funded by the State Government of Western Australia and the Joint Venture Partners.  SJT is a Western Australian PremierÕs Fellow, funded by the State Government of Western Australia.  The authors gratefully acknowledge the assistance of Jan Wagner in the development of phase calibration tone extraction algorithms.

\bibliographystyle{apj}
\bibliography{deller_thesis}

\end{document}